\documentclass[11pt,twoside]{article}
\usepackage{asp2014}

\aspSuppressVolSlug
\resetcounters

\def\ls{\vskip 12.045pt}
\newcommand\starlight{{\sc starlight}}                  
\newcommand\ageL{$\langle {\rm log}\,age\rangle _L$}
\newcommand\logZM{$\langle {\rm log}\,Z_\star\rangle _M$}
\newcommand\ageM{$\langle {\rm log}\,age\rangle _M$}

\begin{document}

\title{The CALIFA survey across the Hubble sequence: How galaxies grow their bulges and disks}
\author{R. M. Gonz\'alez Delgado,$^1$ R. Garc\'{\i}a-Benito,$^1$ E. P\'erez,$^1$ 
R. Cid Fernandes,$^2$ A. L.\ de Amorim,$^2$ C. Cortijo-Ferrero, $^1$ E. A. D. Lacerda,$^{1,2}$ R. L\'opez Fern\'andez,$^1$ 
N. Vale-Asari,$^2$  S. S\'anchez,$^3$ and CALIFA collaboration$^4$
\affil{$^1$Instituto de Astrof\'{\i}sica de Andaluc\'{\i}a (CSIC), P.O. Box 3004, 18080 Granada, Spain; \email{rosa@iaa.es}}
\affil{$^2$Departamento de F\'{\i}sica, Universidade Federal de Santa Catarina, P.O. Box 476, 88040-900, Florian\'opolis, SC, Brazil}
\affil{$^3$Instituto de Astronom\'\i a,Universidad Nacional Auton\'oma
de Mexico, A.P. 70-264, 04510, M\'exico,D.F.}
\affil{$^4$http://califa.caha.es}}

\paperauthor{Sample~Author1}{Author1Email@email.edu}{ORCID_Or_Blank}{Author1 Institution}{Author1 Department}{City}{State/Province}{Postal Code}{Country}
\paperauthor{Sample~Author2}{Author2Email@email.edu}{ORCID_Or_Blank}{Author2 Institution}{Author2 Department}{City}{State/Province}{Postal Code}{Country}
\paperauthor{Sample~Author3}{Author3Email@email.edu}{ORCID_Or_Blank}{Author3 Institution}{Author3 Department}{City}{State/Province}{Postal Code}{Country}

\begin{abstract}
We characterize in detail the radial structure of the stellar population properties of 300 galaxies in the nearby universe, observed with integral field spectroscopy in the CALIFA survey. The sample covers a wide range of Hubble types, from spheroidal to spiral galaxies,  ranging in stellar masses  from $M_\star \sim 10^9$ to $7 \times 10^{11}$ $M_\odot$. We derive the stellar mass surface density ($\mu_\star$),  light-weighted and mass-weighted ages (\ageL, \ageM), and mass-weighted metallicity (\logZM), applying the spectral synthesis technique. We study the mean trends with galaxy stellar mass, $M_\star$, and morphology (E, S0, Sa, Sb, Sbc, Sc and Sd). We confirm that more massive galaxies are more compact, older, more metal rich, and less reddened by dust. Additionally, we find that these trends are preserved spatially with the radial distance to the nucleus. Deviations from these relations appear correlated with Hubble type: earlier types are more compact, older, and more metal rich for a given M$_\star$, which evidences that quenching is related to morphology, but not driven by mass. 
\end{abstract}

\section{Introduction}

Galaxies are a complex mix of stars,  gas, dust, and dark matter, distributed in different components (bulge, disk, and halo) whose present day structure and dynamics  are intimately linked to their assembly and evolution over the history of the Universe. 
Observationally, it has been found that galaxy properties such as color, mass, surface brightness, luminosity, and gas fraction are correlated with Hubble type \citep{R94}, suggesting that the Hubble sequence somehow reflects possible paths for galaxy formation and evolution. However, the processes structuring galaxies along the Hubble sequence are still poorly understood.

Integral Field Spectroscopy (IFS) enables a leap forward for understanding the galaxy formation processes, providing 3D information (2D spatial + 1D spectral) on galaxies. CALIFA (Calar Alto Legacy Integral Field Area) is our ongoing IFS survey of 600 nearby galaxies at the 3.5m at Calar Alto  \citep{sanchez12, RGB15}. This survey is unique to advance in these issues  not only because of its ability to provide spectral and spatial information, but also because it includes a large homogeneous sample of galaxies across the Hubble sequence. 

Previous works have used the first $\sim$100 datacubes of the survey to derive spatially resolved  stellar population properties by means of full spectral fitting techniques.
We have obtained that: 
1) Massive galaxies grow their stellar mass inside-out. The signal of downsizing is shown to be spatially preserved, with both inner and outer regions growing faster for more massive galaxies. The relative growth rate of the spheroidal component (nucleus and inner galaxy), which peaked 5--7 Gyr ago, shows a maximum at a critical stellar mass $M_\star \sim 7 \times 10^{10} M_\odot$ \citep{perez13}. 
2)  Global and local relations between stellar mass, stellar mass surface density and stellar metallicity relation were investigated, along with their evolution (as derived from our fossil record analysis). In disks, the stellar mass surface density regulates the ages and the metallicity. In spheroids, the galaxy stellar mass dominates the physics of star formation and chemical enrichment \citep{RGD14a,RGD14b}.  
3) In terms of integrated versus spatially resolved properties, the stellar population properties are well represented by their values at 1 HLR \citep{RGD14a,RGD14b}.

The goal of this contribution is: a) to characterize in detail the radial structure of stellar population properties of galaxies in the local universe; b) to find out how these properties are correlated with  Hubble type,  and if the Hubble sequence is a scheme to organize galaxies by mass and age, and/or  mass and metallicity; c)  To establish observational constraints to galaxy formation models via the radial distributions and gradients of  stellar populations for disk and bulge dominated galaxies.

\section{Results}

{\it The sample:} comprises 300 CALIFA galaxies that are grouped into 7 morphology bins. They provide a fair representation of the CALIFA survey as a whole, with 41 E, 32 S0, 51 Sa, 53 Sb, 58 Sbc, 50 Sc, and 15 Sd. 

{\it The method:} We apply the fossil record method based on the spectral synthesis technique to recover the stellar population properties. We extract stellar population properties from  the data-cubes, as explained in \citet{Cid2013,Cid2014}. In short, we analyze the data with the \starlight\ code \citep{Cid2005}, which fits an observed spectrum in terms of a model built by a non-parametric linear combination of $N_\star$ Simple Stellar Populations (SSPs) from a base spanning different ages ($t$) and metallicities ($Z$). We use a set that combines the SSP spectra provided by \citet{vazdekis10} and  \citet{RGD05}. The IMF is Salpeter.

{\it Galaxy stellar mass, $M_\star$:}  
The masses range from $7\times 10^8$ to $7\times 10^{11} M_\odot$.  As for the general galaxy population, mass is well correlated with  Hubble type, decreasing from early to late types. High  bulge-to-disk ratios (E, S0, Sa) are the most massive ones ($\geq 10^{11} M_\odot$), while galaxies with small bulges (Sc--Sd) have  $M_\star \leq  10^{10}  M_\odot$. The average $\log M_\star (M_\odot)$ is 11.4, 11.1, 11.0, 10.9, 10.7, 10.1, and 9.5 for E, S0, Sa, Sb, Sbc, Sc, and Sd, respectively. The dispersion is  typically 0.3 dex, except for Sc galaxies, that have a dispersion of $\sim 0.5$ dex.

{\it Galaxy size and the HMR/HLR:} 
 From the  spatially resolved SFH and extinction maps we obtain that the ratio of the radius that contains half of the mass (HMR) with respect to the radius that contains half of the light (HLR) of  HMR/HLR $\sim 0.8$. Galaxies are therefore typically $20 \%$ smaller in mass than they appear in optical light. The distribution along the Hubble sequence shows a clear trend. Sa-Sb-Sbc have the lowest  ratios, due to the fact that they show a very prominent old bulge which have similar central properties to the spheroidal components of S0 and E, but a blue and extended disc which contributes to the light in the range Sa to Sbc.
 
 \begin{figure}
\includegraphics[width=0.5\textwidth]{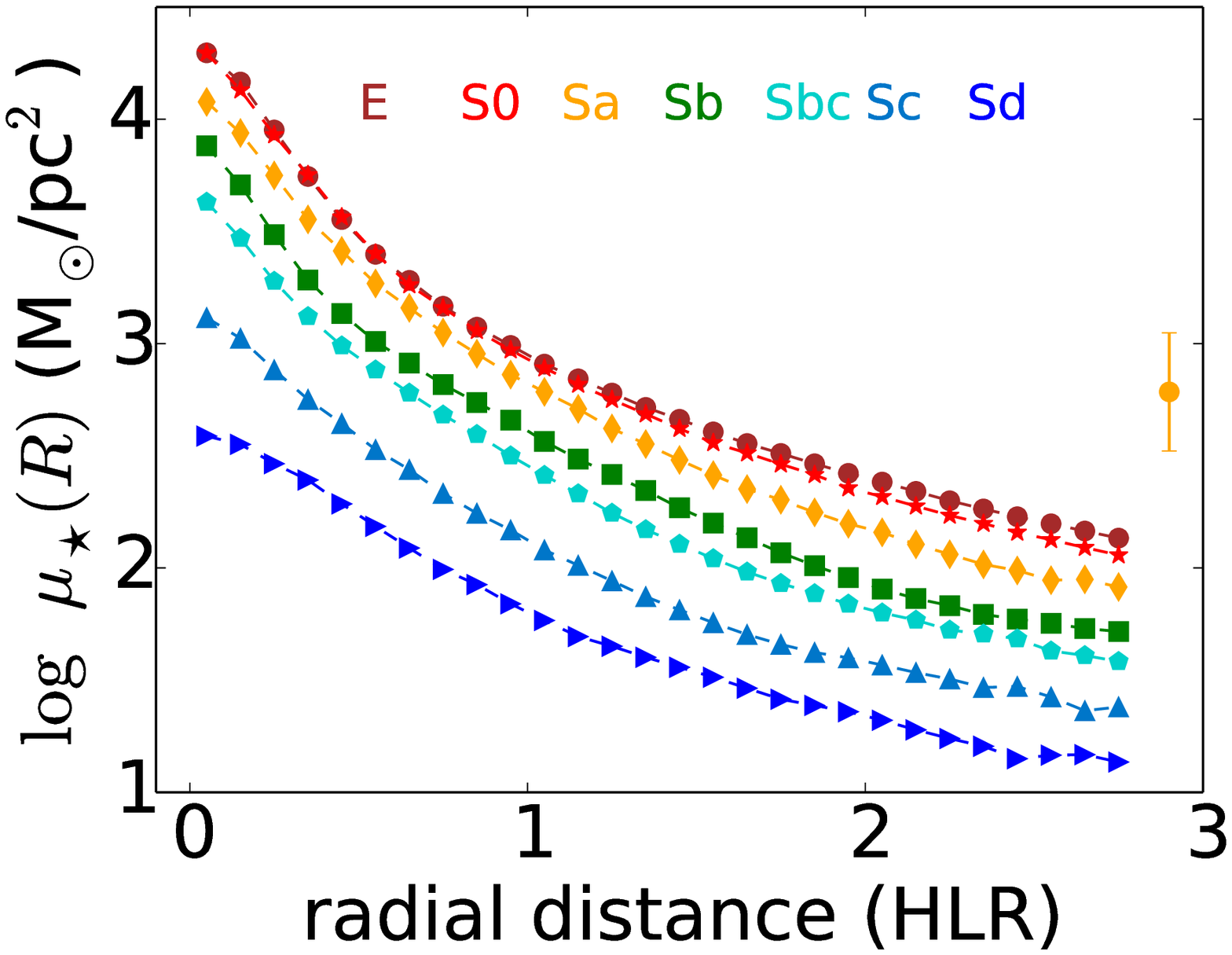}
\includegraphics[width=0.5\textwidth]{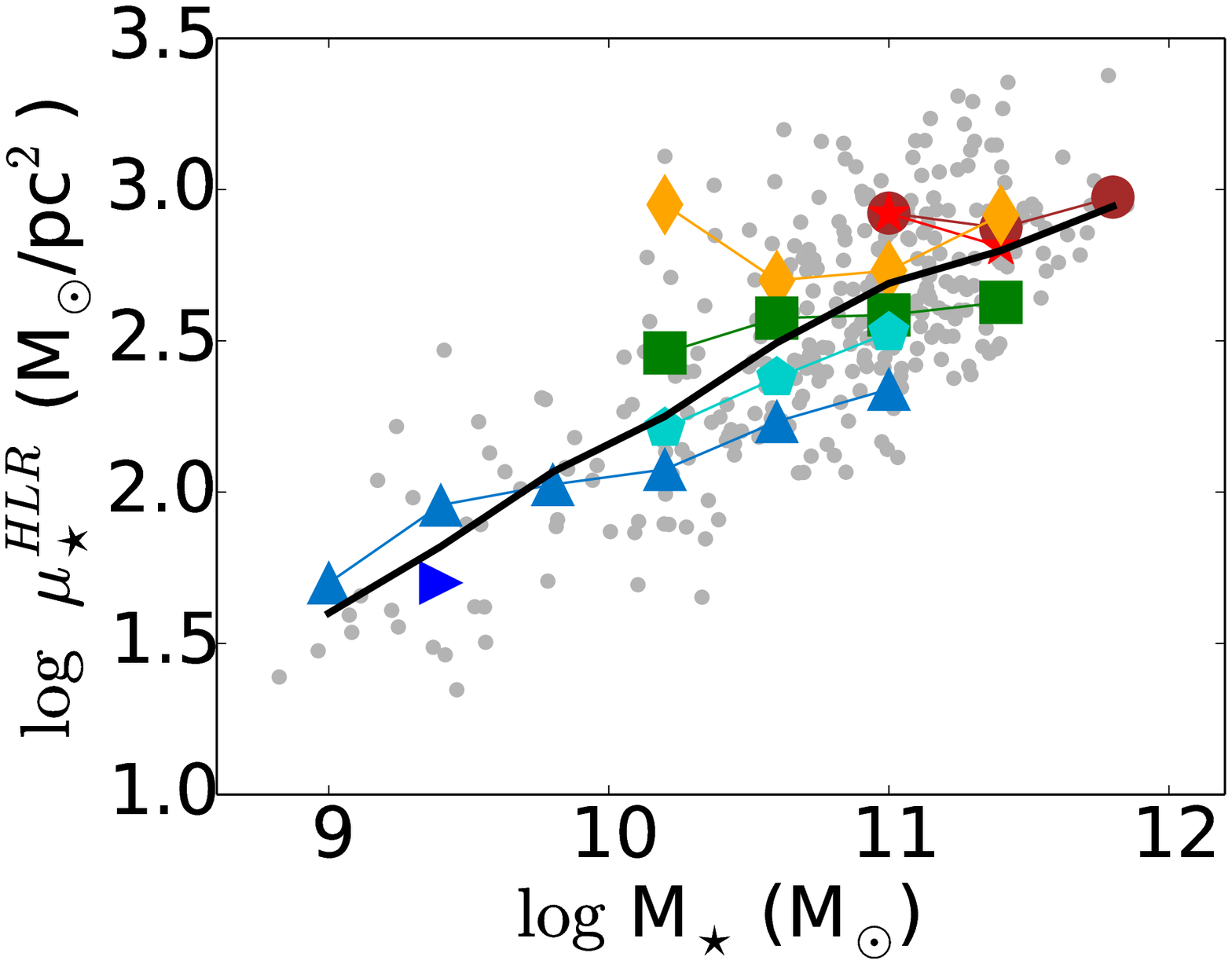}
\caption{$\log \mu_\star$ measured at 1 HLR (right panel) and the radial profiles  for each Hubble type (left panel). Symbols are: circles (E), stars (S0), diamonds (Sa), squares (Sb), pentagons (Sbc), triangles (Sc), triangles right (Sd). }
\end{figure}

{\it Radial profiles: Stellar mass surface density.} $\log \mu_\star (R)$ shows declining profiles that scale with morphology and with $M_\star$; this behavior is preserved at any given distance. At constant $M_\star$, $\log \mu_\star (R)$ is higher in early type than in late type spirals. E's and S0's show equal $\log \mu_\star (R)$ profiles, independently of $M_\star$. The inner gradient, $\bigtriangledown_{in} \log \mu_\star$, correlates  with Hubble type. 
The negative gradients steepen from late type spirals to spheroids, as well as with galaxy total mass in galaxies with $M_\star \leq$ 10$^{11}$ $M_\odot$. At a constant $M_\star$, $\bigtriangledown_{in} \log \mu_\star$ steepens with  morphology, with E's and S0's having the steepest gradients. These results indicate that morphology, and not only $M_\star$, plays a relevant role in defining $\mu_\star$, and the $\mu_\star$$-$$M_\star$ relation (see Fig. 1).

\begin{figure}
\includegraphics[width=0.5\textwidth]{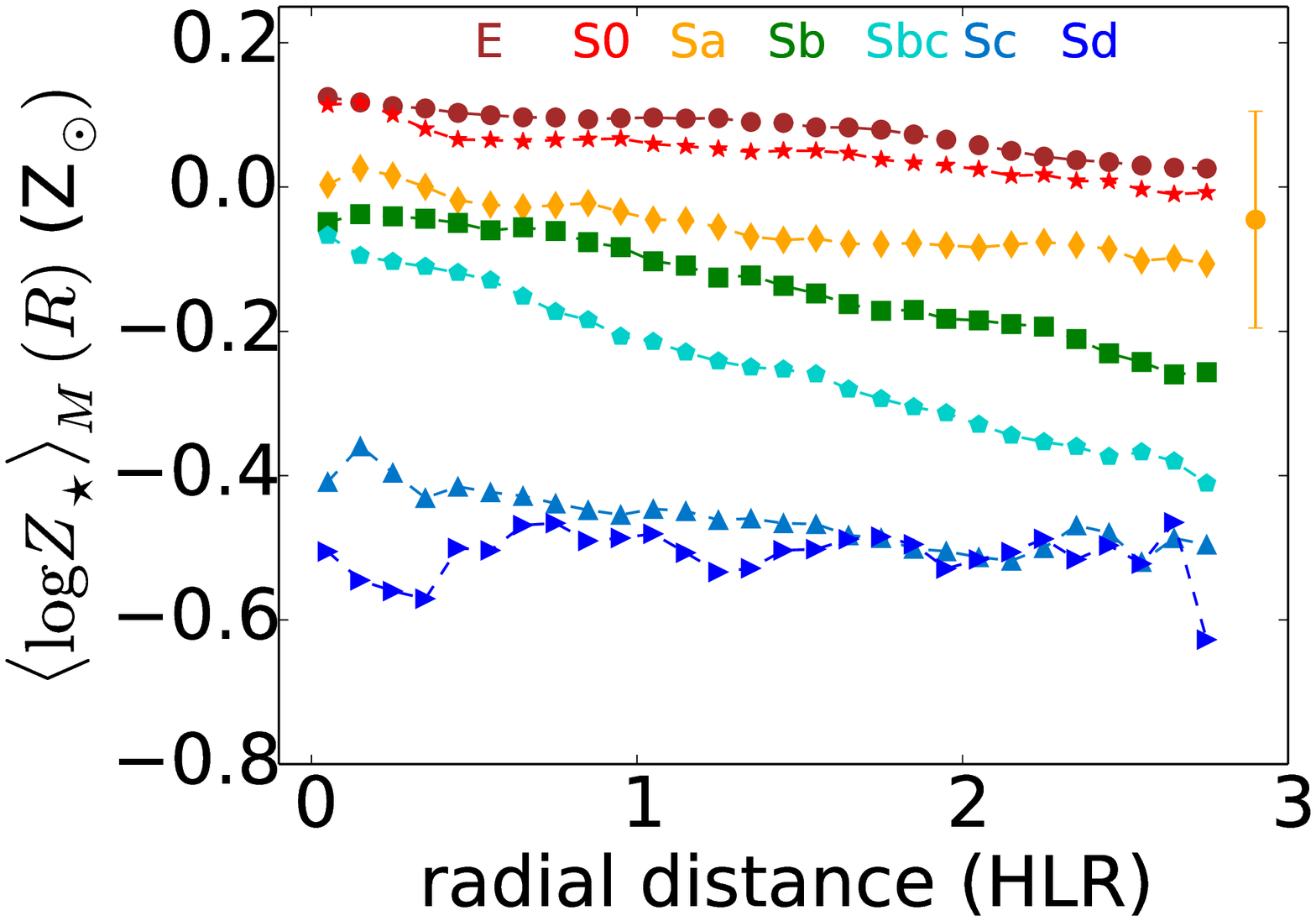}
\includegraphics[width=0.5\textwidth]{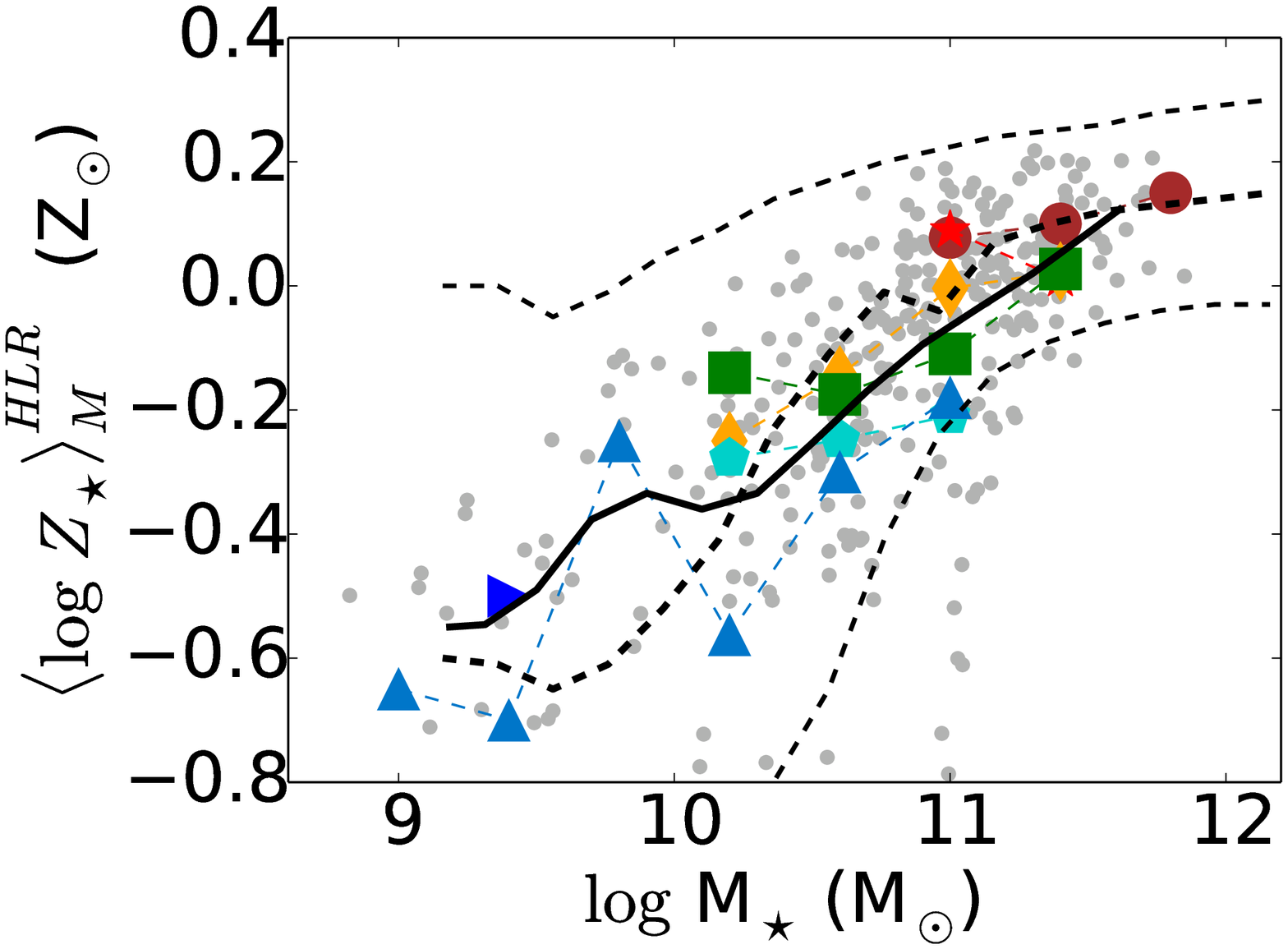}
\caption{As Fig.1 for \logZM\ . Dashed lines show the mass-metallicity relation obtained for SDSS galaxies by  Gallazzi et al. 2005}
\end{figure}

{\it Radial profiles: Stellar metallicity.} \logZM(R) shows mildly decreasing profiles for most  Hubble types, except Sd's that show little, if any, radial dependence. Milky Way like galaxies (Sbc) stand out as the ones with the steepest radial profiles.  \logZM(R) scales with $M_\star$ in a similar way as it does with  morphology. This can be understood as a consequence of the global mass metallicity relation --a primary dependence of the metallicity with $M_\star$. The metallicity gradients are negative but shallow on average, with $\bigtriangledown_{in}$\logZM\  $\sim -0.1$ dex/HLR, and show a small dependence with $M_\star$ up to M$_\star \sim$ 10$^{11}$ $M_\odot$, steepening with increasing mass. Above 10$^{11}\, M_\odot$ (Sa's, S0's and E's) they have similar metallicity gradient. The dispersion in the  $\bigtriangledown_{in}$\logZM\ $-$ $M_\star$ relation is significant and a trend with  morphology is seen, in the sense that, for a given mass, intermediate type spirals are the ones with steeper gradients (see Fig. 2). 

\begin{figure}
\includegraphics[width=0.5\textwidth]{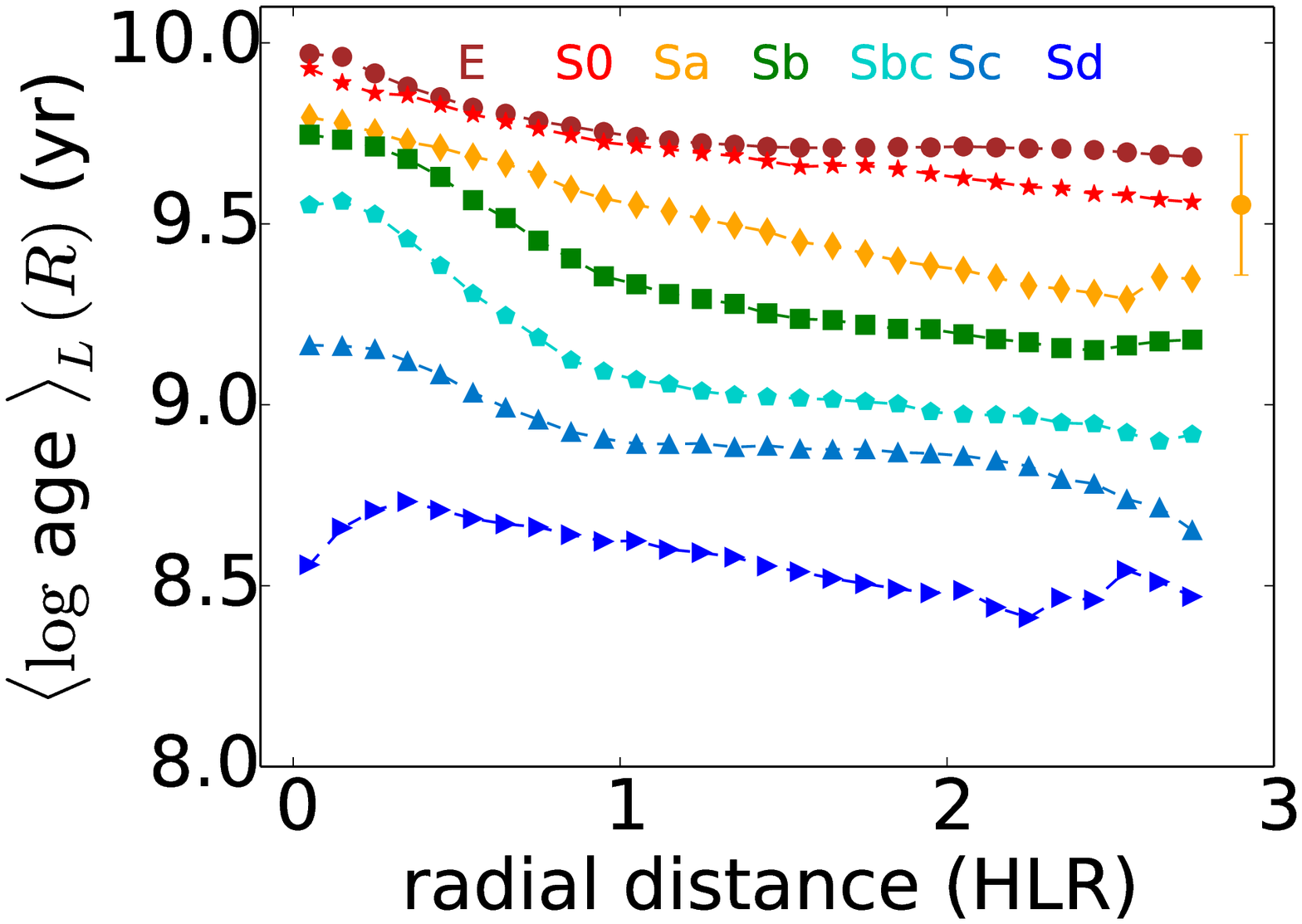}
\includegraphics[width=0.5\textwidth]{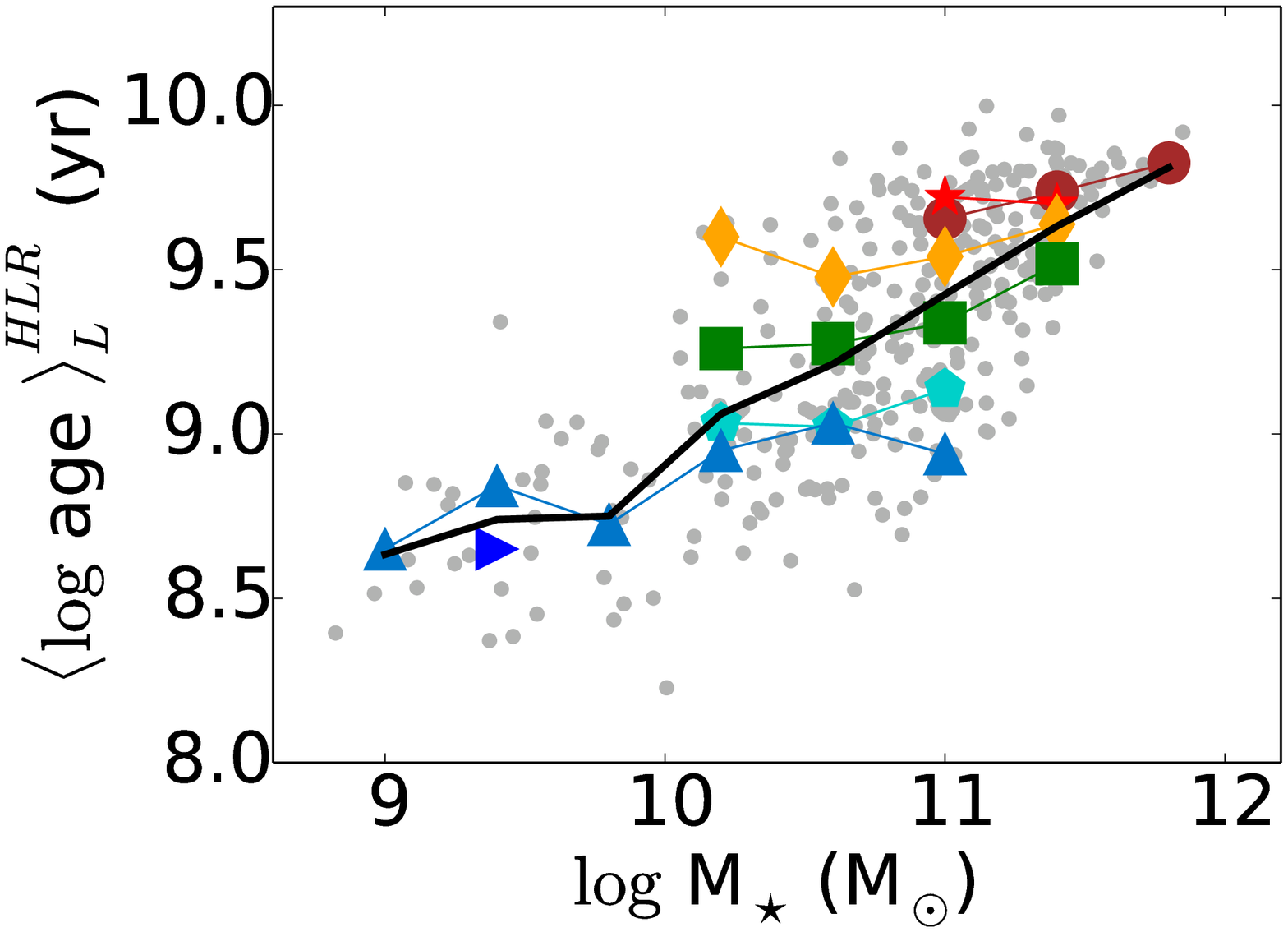}
\caption{As Fig.1 for \ageL\ }
\end{figure}

{\it Radial Profiles: Stellar ages}. \ageL(R)  shows declining profiles that scale with morphology; this behavior is preserved at any given distance.  Early type spirals are always older than late spirals. E's and S0's, although older than spirals, have both similar \ageL(R) profiles, indicating that these galaxies have similar star formation histories. The more massive galaxies are also the older ones; this ``downsizing" behavior is always preserved at any given distance. The negative $\bigtriangledown_{in}$\ageL\  depends on  Hubble type in different ways: steeper from E and S0 to Sbc, and shallower from Sbc to Sd. Thus, Milky Way like galaxies have the steepest age gradient. A $\bigtriangledown_{in}$\ageL $-M_\star$ relation exists, increasing the gradient from the low mass galaxies (which have roughly flat profiles) up to about 10$^{11} M_\odot$, at this point the trend reverses and $\bigtriangledown_{in}$\ageL\ decreases with increasing $M_\star$. However, the dispersion in the $\bigtriangledown_{in}$\ageL $-M_\star$ relation and \ageL$^{HLR}$$-M_\star$ is significant and it is strongly related with the morphology.  Even more, the dispersion of the \ageL(R) profiles of galaxies of equal mass is significant and larger than between the \ageL(R) profiles of galaxies of different $M_\star$ but the same Hubble type. Thus, the SFHs and their radial variations are modulated primarily by the Hubble type, with mass playing a secondary role (see Fig. 3).

\section{Conclusions}
\begin{enumerate}

\ls\item{Evidence in favor of the inside-out growth of galaxies is found in the negative radial stellar age gradients. Metallicity gradients and the fact that galaxies are more compact in mass than in light also support this scenario.}

\ls\item{The mean stellar ages of disks and bulges are correlated, with disks covering a large range of ages, and late type spirals hosting the younger disks.
The bulges of S0 and early type spirals are old and metal rich as the core of E's.
They formed by similar processes, through mergers. Later type spirals, however,  have younger bulges, and larger contribution from secular evolution are expected. Disks are younger and more metal poor than bulges, as an indicative of the inside-out formation scenario of these galaxies.}

\ls\item{ It is the Hubble type, not $M_\star$, that drives differences in the  galaxy averaged age, and radial age gradients. These results indicate that the SFH and their radial variations are modulated primarily by galaxy morphology, and only secondarily by the galaxy mass.  This suggest that galaxies are  morphologically quenched, and that the shutdown of  star formation occurs outwards and earlier in galaxies with a large spheroid than in galaxies of later Hubble type. }

\end{enumerate}

Thanks to the uniqueness of the CALIFA data in terms of  spatial coverage and resolution,  large sample spanning all morphological types, and homogeneity and quality of the spectral analysis, we are able to characterize the radial structure of the stellar population properties of galaxies in the local universe. The results show that the Hubble sequence is a useful scheme to organize galaxies by their spatially resolved stellar density, ages, and metallicity. Stellar mass, although responsible for setting the average stellar population properties in galaxies, it is less responsible of the quenching processes. Morphology is, however, more strongly connected with the shut down of the star formation activity in the bulges and disks of galaxies.

\acknowledgements This contribution is based on data obtained by the CALIFA survey (http://califa.caha.es), funded by the Spanish MINECO grants ICTS-2009-10, AYA2010-15081, Junta de Andaluc\'\i a FQ1580, and the CAHA operated jointly by the Max-Planck IfA and the IAA (CSIC). The CALIFA Collaboration thanks the CAHA staff for the dedication to this project. Support from CNPq (Brazil) through Programa Ci\^encia sem Fronteiras (401452/2012-3) is duly acknowledged.


\begin{thebibliography}{}


\bibitem[\protect\citeauthoryear{Cid Fernandes et al.}{2005}]{Cid2005} Cid Fernandes, R.,  Mateus, A.,  Sodr\'e, L.,  et al.\ 2005, \mnras, 358, 363

\bibitem[\protect\citeauthoryear{Cid Fernandes et al.}{2013}]{Cid2013} Cid Fernandes, R., P\'erez, E., Garc\'\i a Benito, R., et al.\ 2013,  \aap, 557, 86

\bibitem[\protect\citeauthoryear{Cid Fernandes et al.}{2014}]{Cid2014} Cid Fernandes, R., Gonz\'alez Delgado, R.M., P\'erez, E., et al.\ 2014, \aap, 561, 130


\bibitem[\protect\citeauthoryear{gallazzi}{2015}]{gallazzi15} Gallazzi, A.,  Charlot, S., Brinchmann, J.,  White, S.D.M., Tremonti, C.A. 2015, \mnras, 362, 41 


\bibitem[\protect\citeauthoryear{Garc\'\i a-Benito et al.}{2015}]{RGB15} Garc{\'{\i}}a-Benito, R., Zibetti, S., S{\'a}nchez, S.~F., et al.\ 2015, \aap, 576, A135 


\bibitem[\protect\citeauthoryear{Gonz\'alez Delgado et al.}{2005}]{RGD05} Gonz\' alez Delgado, R. M., Cervi\~no, M., Martins, et al.\ 2005, \mnras, 357, 945

\bibitem[\protect\citeauthoryear{Gonz\'alez Delgado et al.}{2014a}]{RGD14a} Gonz\' alez Delgado, R. M., P\'erez, E., Cid Fernandes, R., et al.\ 2014a, \aap, 562, 47 

\bibitem[\protect\citeauthoryear{Gonz\'alez Delgado et al.}{2014b}]{RGD14b} Gonz\' alez Delgado, R. M., Cid Fernandes, R., Garc\'\i a-Benito, R., et al.\ 2014b, \apjl, 791, L16



\bibitem[\protect\citeauthoryear{P\'erez et al.}{2013}]{perez13} P\'erez, E., Cid Fernandes, R., Gonz\'alez Delgado, R. M., et al.\ 2013, \apjl, 764, 1L


\bibitem[Roberts \& Haynes(1994)]{R94} Roberts, M.~S., \& Haynes, M.~P.\ 1994, \araa, 32, 115 

\bibitem[\protect\citeauthoryear{S\'anchez, S.F.  et al.}{2012}]{sanchez12} S\'anchez, S. F., Kennicutt, R. C., Gil de Paz, A., et al.\ 2012, \aap, 538, 8

\bibitem[\protect\citeauthoryear{Vazdekis  et al.}{2010}]{vazdekis10} Vazdekis, A., S\'anchez-Bl\'azquez, P., Falc\'on-Barroso, J., et al.\ 2010, \mnras, 404, 1639

\end{thebibliography}


\end{document}